\documentclass[floats,floatfix,showpacs,amssymb,prd,nofootinbib,superscriptaddress,aps,notitlepage,reprint]{revtex4-1}
\pdfoutput=1
\usepackage{graphicx}
\usepackage[utf8,latin1]{inputenc}
\usepackage[english]{babel}
\usepackage{graphicx, epsfig, amssymb} 
\usepackage{amsmath, amsfonts}
\usepackage{bm} 
\usepackage[normalem]{ulem} 
\usepackage{soul}
\usepackage[linktocpage]{hyperref}
\usepackage[caption=false]{subfig}
\usepackage[usenames]{color}

\def\be{\begin{equation}}
\def\ee{\end{equation}}

\newcommand{\diff}{\textrm{d}}

\newcommand{\bea}{\begin{eqnarray}}
\newcommand{\eea}{\end{eqnarray}}
\newcommand{\ben}{\begin{enumerate}}
\newcommand{\een}{\end{enumerate}}
\newcommand{\bi}{\begin{itemize}}
\newcommand{\ei}{\end{itemize}}

\def\ga{\mathrel{\raise.3ex\hbox{$>$\kern-.75em\lower1ex\hbox{$\sim$}}}}
	\def\la{\mathrel{\raise.3ex\hbox{$<$\kern-.75em\lower1ex\hbox{$\sim$}}}}

\def\be{\begin{equation}}
\def\ee{\end{equation}}

\def\I_M{{I_{\scriptscriptstyle M\times M}}}

\def\be{\begin{equation}}
\def\ee{\end{equation}}
\def\bea{\begin{eqnarray}}
\def\eea{\end{eqnarray}}
\newcommand{\beq}{\begin{eqnarray}}
\newcommand{\eeq}{\end{eqnarray}}

\newcommand{\beqal}{\begin{eqnarray}\label}
\newcommand{\none}{\end{eqnarray}}
\newcommand{\beqa}{\begin{eqnarray}}
\newcommand{\eeqa}{\end{eqnarray}}

\begin{document}
\title{\large Gravitational waves emitted by a particle rotating around a Schwarzschild black hole: A semiclassical approach}

\author{Rafael P. Bernar}\email{rafael.bernar@icen.ufpa.br}
\affiliation{Faculdade de F\'{\i}sica, Universidade
Federal do Par\'a, 66075-110, Bel\'em, Par\'a, Brazil.}

\author{Lu\'is C. B. Crispino}\email{crispino@ufpa.br}
\affiliation{Faculdade de F\'{\i}sica, Universidade
Federal do Par\'a, 66075-110, Bel\'em, Par\'a, Brazil.}

\author{Atsushi Higuchi}\email{atsushi.higuchi@york.ac.uk}
\affiliation{Department of Mathematics, University of York, YO10 5DD, Heslington, York, United Kingdom.}

\begin{abstract}	
We analyze the gravitational radiation emitted from a particle in circular motion around a Schwarzschild black hole using the framework of quantum field theory in curved spacetime at tree level. The gravitational perturbations are written in a gauge-invariant formalism for spherically symmetric spacetimes. We discuss the results, comparing them to the radiation emitted by a particle when it is assumed to be orbiting a massive object due to a Newtonian force in flat spacetime. 
\end{abstract}

\pacs{
04.60.-m, 
04.62.+v, 
04.50.-h, 
04.25.Nx, 
04.60.Gw, 
11.25.Db  
}

\date{\today}

\maketitle


\section{Introduction}

Black holes are among the most important predictions of General Relativity (GR). Several observations indicate the presence of supermassive black holes in the center of nearly all large galaxies \cite{1998Natur.395A..14R,doi:10.1146/annurev-astro-082708-101811}. In addition, there is strong evidence for stellar-mass black holes having an influence on other stars in binary systems \cite{lewin1997x}, emitting X-rays through accretion (see Ref.~\cite{sciama} for a review on observational evidence of stellar-mass and supermassive black holes). Moreover, black holes are believed to play an important part in powerful astrophysical processes, such as gamma-ray bursts \cite{gammaraybursts}. The recent detections of gravitational waves~\cite{gw140915, ref:GW151226} emitted by binary black hole mergers make the study of black holes and radiation-emission scenarios even more appealing, particularly the emission of gravitational waves. Binary black hole systems can provide settings in which the extreme curvature of the black hole generates remarkable signatures which can, in principle, be experimentally detected. It is also interesting to study gravitational radiation emitted by a relatively small object which can be approximated by a point particle in circular orbit around a black hole, in highly relativistic motion, the so-called geodesic synchcrotron radiation scenario. The possibility of this mechanism for gravitational synchrotron radiation was raised in Refs.~\cite{PhysRevLett.28.994, misnergsr}. While studying the scalar radiation emitted by a point source in circular geodesic motion around a Schwarzschild black hole, it was argued that gravitational radiation emitted by the source would be mostly of the synchrotron type, which has
frequencies much higher than the angular frequency of the orbit and radiation distributed in narrow angles. This was further investigated in Ref.~\cite{chitre} where the authors computed the high-frequency spectra of electromagnetic and gravitational radiation for a particle orbiting a Schwarzschild black hole. 
In Refs.~\cite{davisruffini, PhysRevD.7.1002}, using the Regge-Wheeler formalism~\cite{regge-wheeler1957,PhysRevD.1.3514,zerilliprl} and Green's function techniques~\cite{zerilliprd,Davis:1971pa,Davis:1971gg,Davis:1972ud}, it was shown that the spectrum of gravitational radiation from a point particle around a Schwarzschild black hole is much broader than the scalar or electromagnetic ones, at least for high-$l$ multipole modes. For stable orbits, a full analysis and numerical computations were done in Refs.~\cite{poissongsr1,poisson2}. 

The framework of Quantum Field Theory (QFT) in curved spacetimes \cite{birrelldavies, parkertoms} has been used at tree level to compute the (massless) scalar radiation of a point source in circular orbit around Schwarzschild \cite{crispino1,crispinoprd77}, Reissner-Nordstr\"om \cite{silva} and Kerr black holes \cite{crispinomacedo}. The case of massive scalar radiation from a point source orbiting a stellar object or a Schwarzschild black hole was analyzed in Ref. \cite{meirafilho}. Electromagnetic radiation from a point charge rotating around an uncharged static black hole was analyzed in Ref. \cite{crispino2}, using the same semiclassical approach. Although this approach is found to give the same results as the classical methods (e.g. using Green's function techniques), it will make the radiative quantum corrections to these results more straightforward. It also allows an alternative interpretation of the radiation processes discussed in this paper. We also find that by regarding the classical fields as quantum particles one can treat several aspects of the radiation phenomena in an unified manner.

In this paper we analyze the gravitational radiation emitted by a point particle in geodesic circular orbit around a Schwarzschild black hole using the framework of QFT in curved spacetimes at tree level. Using numerically obtained solutions, we compute the total emitted power, as well as the power radiated to infinity in both stable and unstable orbits. We also analyze the spectrum of the emitted radiation.


The rest of this paper is organized as follows. In Sec.~\ref{sec:perturbationsschwarz} we present the formalism developed in Refs.~\cite{kis2000,kodamaishibashi} for linear perturbations of the gravitational field around spherically symmetric spacetimes. We give a brief review of the formalism, specializing it to the background of a 4-dimensional Schwarzschild spacetime. In Sec.~\ref{sec:quantumgsr} we present the framework of QFT for linearized gravity in which we will work, applying this framework to the case of a point particle emitting gravitational radiation, in geodesic circular motion around a Schwarzschild black hole. We also obtain numerical results for the emitted power of gravitational radiation. In Sec.~\ref{sec:compareflat} we compare these results to an analogous case in flat spacetime, namely the radiation emitted by a particle orbiting a Newtonian massive object. We conclude this paper with some remarks in 
Sec.~\ref{sec:finalremarks}. We present the derivation of the normalization factor of one type of the modes, the scalar-type modes, in Appendix~\ref{app:scalarinner}. Throughout this paper we use the metric  signature $-+++$ and natural units such that $G=c=\hbar=1$.

\section{Gravitational perturbations in Schwarzschild spacetime}
\label{sec:perturbationsschwarz}
In this section we present a brief review of the formalism developed in Refs. \cite{kis2000,kodamaishibashi}. By expanding suitably defined gauge-invariant quantities in terms of harmonic tensors, the perturbed Einstein's equations are reduced to a set of self-adjoint ordinary differential equations, one for each type of perturbation: scalar-, vector- and tensor-type gravitational perturbations. This formalism can be used for background spacetimes in arbitrary dimensions with some special isometries. Here we restrict this formalism to Schwarzschild spacetime in 3+1 dimensions.
\subsection{Background Schwarzschild Spacetime} \label{subsec:background}
We work in the background spacetime of a chargeless nonrotating black hole of mass $M$, described by the line element:
\beqa
\diff s^2&=&g_{\mu\nu}\diff x^{\mu}\diff x^{\nu} \nonumber \\
&=&-f(r)\diff t^2+\frac{\diff r^2}{f(r)}+r^2(\diff\theta^2 + \sin ^2 \theta \diff\varphi^2), \label{schwarzschildST}
\eeqa
where
\be
f(r)=1-\frac{2M}{r}.
\ee
It is useful for us to define the line element of the orbit spacetime:
\be
\diff s^2_{orb}=g_{ab}\diff x^{a}\diff x^{b}=-f(r)\diff t^2+\frac{\diff r^2}{f(r)}. \label{orbit}
\ee
As we will see in Sec.~\ref{subsec:scalarandvectorpert} it is basically in the orbit spacetime that the dynamical equations for perturbations have to be solved.
It is also useful to define the line element of the two-sphere $S^2$:
\be
\diff\sigma^2=\gamma_{ij}\diff x^{i}\diff x^{j}=\diff\theta^2 + \sin^2 \theta \diff\varphi^2. \label{2-sphere}
\ee
The definitions (\ref{schwarzschildST}), (\ref{orbit}) and (\ref{2-sphere}) above establish part of the notation used. Greek letters are used for spacetime indices running from 0 to 3, the first letters from the Latin alphabet are used for the $t$ and $r$ components and letters $i,j,k,...$ are used for the $\theta$ and $\varphi$ components. Covariant derivatives and Christoffel symbols corresponding to $\diff s^2$, $\diff s^2_{orb}$ and $\diff\sigma^2$ are denoted by $\nabla_{\mu}$, $\Gamma^{\alpha}_{\mu\nu}$; $D_{a}$, $\Gamma^{a}_{bc}$; and $\hat{D}_{i}$, $\hat{\Gamma}^{i}_{jk}$ respectively.

\subsection{Scalar-type and vector-type perturbations} \label{subsec:scalarandvectorpert}
Gravitational perturbations of the scalar-type are defined as the metric perturbations whose angular dependence is described by the scalar spherical harmonics $Y^{lm}(\theta,\varphi)$, which satisfy
\be
\left(\hat{\Delta}_2 + k_S^2\right)Y^{lm}(\theta,\varphi)=0, \label{scalarsphericaleq}
\ee
with eigenvalues
\be
k_S^2=l(l+1), \hspace{0.5cm} l=0,1,2,...,
\ee
where $\hat{\Delta}_2$ is the Laplace-Beltrami differential operator on $S^2$, namely
\be
\hat{\Delta}_2 \equiv \frac{1}{\sin\theta}\frac{\partial}{\partial \theta}\left(\sin\theta\frac{\partial}{\partial \theta}\right)+\frac{1}{\sin^2\theta}\frac{\partial^2}{\partial \varphi^2}.
\ee
The solutions to Eq.~(\ref{scalarsphericaleq}) are given by
\be
Y^{lm}(\theta,\varphi)=C^{lm}P^{l}_{m}(\cos\theta)e^{im\varphi}.
\ee
The normalization constants are \cite{cohen-tannoudji}:
\be
C^{lm}=\sqrt{\frac{(2l+1)}{4\pi}\frac{(l-m)!}{(l+m)!}}.
\ee
The scalar-type metric perturbation modes $h_{\mu\nu}^{(S;lm)}$ can be written as follows \cite{kodamaishibashi}
\beqa
h_{ab}^{(S;lm)}&=&f_{ab}^{(S;l)}Y^{lm},\\
h_{ai}^{(S;lm)}&=&rf_{a}^{(S;l)}\mathbb{S}^{(lm)}_i,\\
h_{ij}^{(S;lm)}&=&2r^2\left(H_L^{S;l}\gamma_{ij}Y^{lm}+H_T^{S;l}\mathbb{S}^{(lm)}_{ij}\right),
\eeqa
where
\beqa
\mathbb{S}^{(lm)}_i&=&-\frac{1}{k_S}\hat{D}_{i}Y^{lm},\\
\mathbb{S}^{(lm)}_{ij}&=&\frac{1}{k_S^2}\hat{D}_{i}\hat{D}_{j}Y^{lm}+\frac{1}{2}\gamma_{ij}Y^{lm}.
\eeqa
The quantities $f_{ab}^{(S;l)}$, $f_{a}^{(S;l)}$, $H_L^{S;l}$ and $H_T^{S;l}$ are gauge-dependent quantities and functions of $t$ and $r$ only.

Gauge-invariant quantities can be defined for $l\geq2$ and written in terms of a master variable $\Phi^{S}_{l}(t,r)$, which satisfies the following wave equation, resulting from the perturbed Einstein's equations:
\beq
\Box \Phi^{S}_{l}(t,r)-\frac{V_S(r)}{f(r)}\Phi^{S}_{l}(t,r)=0, \label{waveeqscalar}
\eeq
with the Zerilli effective potential \cite{zerilliprl}:
\beqa
V_S(r)=f(r)\frac{2\lambda^2(\lambda+1)r^3+6\lambda^2Mr^2+18\lambda M^2r+18M^3}{r^3(\lambda r+3M)^2}, \nonumber\\ \label{scalarpotential}
\eeqa
where
\beq
\lambda=\frac{1}{2}(l-1)(l+2).
\eeq
The $\Box$ is the d'Alembert operator in the orbit spacetime, namely
\beq
\Box \equiv -f(r)^{-1}\frac{\partial^2}{\partial t^2}+\frac{\partial}{\partial r}\left[f(r)\frac{\partial}{\partial r}\right].
\eeq
The derivation of Eq.~(\ref{waveeqscalar}) is highly involved and can be found in Refs. \cite{kodamaishibashi, kis2000}.
Using the same gauge choice as in Refs. \cite{hbc1,ref:hbcproc}, one can write the perturbation modes in terms of the master variable $\Phi^{S}_{l}(t,r)$ as
\beqa
h_{ai}^{(S;lm)}&=&0,\label{scalarmodes1} \\ 
h_{ij}^{(S;lm)}&=&2r^2\gamma_{ij}F^{l}Y^{lm}, \label{scalarmodes2} \\ 
h_{ab}^{(S;lm)}&=&F^{(l)}_{ab}Y^{lm}, \label{scalarmodes3}
\eeqa
with
\beqa
F^{l}=\frac{1}{4r^2H}\left[(H-rf')\Omega^{S}_{l}+2rD^{a}rD_{a}\Omega^{S}_{l}\right], \nonumber\\ \label{F}
\eeqa
\beq
F^{(l)}_{ab}=\frac{1}{H}\left(D_aD_b-\frac{1}{2}g_{ab}\Box\right)\Omega^{S}_{l}, \label{Fab}
\eeq
\beq
H=2\left(\lambda+\frac{3M}{r}\right), \label{eq:hfunction}
\eeq
and
\beq
\Omega^{S}_{l}=rH\Phi^{S}_{l}.
\eeq

The mode with $l=0$ cannot be described in terms of the master variable of Eq.~(\ref{waveeqscalar}). However, it is a spherically symmetric perturbation, which, by Birkhoff's theorem, consists in a shift of the mass parameter of the black hole \cite{zerilliprd}. 
Hence, we will not consider this mode, since it is non-radiative. The $l=1$ modes can always be eliminated by a gauge transformation \cite{kodamaishibashi}.

Gravitational vector-type perturbations are defined as the metric perturbations whose angular dependence is described by vector spherical harmonics satisfying the following equations
\beqa
\left(\hat{\Delta}_2+k_V^2\right)Y^{(lm)}_{i}(\theta,\varphi)&=&0, \label{vectorharmonicseq1}\\
\hat{D}^{j}Y^{(lm)}_{j}(\theta,\varphi)&=&0. \label{vectorharmonicseq2}
\eeqa
The set of eigenvalues takes the form
\beq
k_V^2=l(l+1)-1, \hspace{0.5cm} l=1,2,3,....
\eeq
The solutions to Eqs.~(\ref{vectorharmonicseq1}) and (\ref{vectorharmonicseq2}) on the unit 2-sphere are \cite{regge-wheeler1957,Higuchi:1986wu}
\beq
Y^{(lm)}_{i}(\theta,\varphi)=\frac{\epsilon_{ij}}{\sqrt{l(l+1)}}\hat{\partial}^{j}Y^{lm}(\theta,\varphi).
\eeq
The Levi-Civita tensor on the $S^2$ is defined by
\beqa
\epsilon_{\theta\theta}&=&\epsilon_{\varphi\varphi}=0, \\
\epsilon_{\theta\varphi}&=&-\epsilon_{\varphi\theta}=\sin\theta.
\eeqa
The gravitational perturbation modes of the vector-type $h^{(V;lm)}_{\mu\nu}$ can be written as
\beqa
h_{ab}^{(V;lm)}&=&0, \\
h_{ai}^{(V;lm)}&=&rf_a^{(V;l)}Y^{(lm)}_{i}, \\
h_{ij}^{(V;lm)}&=&2r^2H_T^{V;l}\mathbb{V}^{(lm)}_{ij},
\eeqa
with
\beq
\mathbb{V}^{(lm)}_{ij}=-\frac{1}{2k_V}\left(\hat{D}_{i}Y^{(lm)}_{j}+\hat{D}_{j}Y^{(lm)}_{i}\right).
\eeq
By defining the gauge-invariant quantity for the modes with $l \geq 2$,
\beq
F_{a}^{(V;l)}=f_{a}^{(V;l)}+\frac{r}{k_V}D_{a}H_T^{V;l},
\eeq
we can write it in terms of a master variable $\Phi^{V}_{l}(t,r)$ as
\beq
rF_{a}^{(V;l)}=\epsilon_{ab}D^{b}\left(r\Phi^{V}_{l}\right).
\eeq
The master variable satisfies the following equation:
\beq
\Box \Phi^{V}_{l}(t,r)-\frac{V_V(r)}{f(r)}\Phi^{V}_{l}(t,r)=0, \label{waveeqvector}
\eeq
with the Regge-Wheeler effective potential \cite{regge-wheeler1957}:
\beq
V_V(r)=f(r)\left(\frac{l(l+1)}{r^2}-\frac{6M}{r^3}\right). \label{vecpotential}
\eeq
The $l=1$ vector-type modes correspond to rotational perturbations, i.e. perturbations which give nonzero angular momentum to the background metric~\cite{zerilliprd}. Again, we will not consider these non-radiative modes. 
In a specific gauge choice \cite{hbc1,ref:hbcproc}, one can write the vector-type modes as follows
\beq
h_{ai}^{(V;lm)}=Y^{(lm)}_{i}\epsilon_{ab}D^{b}\left(r\Phi^V_{l}\right), \label{vecmodes}
\eeq 
with all other components vanishing, where $\epsilon_{ab}$ is the Levi-Civita tensor in the orbit spacetime.

In $n+1$ dimensional spacetime with $n\geq 4$, there are also gravitational tensor-type perturbations, whose angular dependence is given by
traceless tensor spherical harmonics.
However,
they
do not exist on $S^2$~\cite{regge-wheeler1957, rubinordonez}. (A concise proof of this fact can be found in Ref.~\cite{Higuchi:1986wu}.) Thus, there are no tensor-type modes for gravitational perturbations in $3+1$ dimensions.


\section{Quantization and geodesic synchrotron radiation}
\label{sec:quantumgsr}
We will consider the case of a test\footnote{The word ``test'' is used here in the sense that the particle does not modify the background metric field.} point particle in circular orbit emitting gravitational waves as it rotates around the black hole. We compute the emitted power using a semiclassical analysis, i.e. by considering the gravitational perturbations as a quantized field in the background Schwarzschild spacetime.

\subsection{Quantization of gravitational perturbations in Schwarzschild spacetime} \label{subsec:quantization}
We quantize the field $h_{\mu\nu}$ in the same manner as in Refs. \cite{hbc1,ref:hbcproc} (see Ref. \cite{Fewster:2012bj} for a more complete description.) The Lagrangian density of free linearized gravity in a background spacetime can be written as:
\beqa
\mathcal{L}&=&\sqrt{-g}\left[\nabla_{\mu}h^{\mu\lambda}\nabla^{\nu}h_{\nu\lambda}-\frac{1}{2}\nabla_{\lambda}h_{\mu\nu}\nabla^{\lambda}h^{\mu\nu}\right. \nonumber \\
&&\left. +\frac{1}{2}\left(\nabla^{\mu}h-2\nabla_{\nu}h^{\mu\nu}\right)\nabla_{\mu}h+R_{\mu\nu\lambda\sigma}h^{\mu\lambda}h^{\nu\sigma} \right], \nonumber \\
\eeqa 
where $h \equiv {h^{\mu}}_{\mu}$.
The conjugate momentum current is given by
\beq
p^{\lambda\mu\nu} &\equiv & \frac{1}{\sqrt{-g}}\frac{\partial \mathcal{L}}{\partial (\nabla_{\lambda}h_{\mu\nu})},
\eeq
thus
\beqa
p^{\lambda\mu\nu}&=& -\nabla^{\lambda}h^{\mu\nu}+g^{\lambda\mu}\left(\nabla_{\kappa}h^{\kappa\nu}-\frac{1}{2}\nabla^{\nu}h\right) \nonumber\\
&&+g^{\lambda\nu}\left(\nabla_{\kappa}h^{\kappa\mu}-\frac{1}{2}\nabla^{\mu}h\right) +g^{\mu\nu}\left(\nabla^{\lambda}h-\nabla_{\kappa}h^{\lambda\kappa}\right). \nonumber \\
\eeqa
Note that we have not yet chosen any gauge condition. For any two solutions to the Euler-Lagrange equations, we define their symplectic product by
\beq
\Omega(h,h')\equiv -\int\limits_{\Sigma}\diff\Sigma n_{\alpha}\left(h_{\mu\nu}p^{\prime\alpha\mu\nu}-p^{\alpha\mu\nu}h'_{\mu\nu}\right), \label{simplecticproduct}
\eeq   
where $p^{\alpha\mu\nu}$ and $p^{\prime\alpha\mu\nu}$ are the conjugate momentum currents of the two solutions $h_{\mu\nu}$ and $h'_{\mu\nu}$, respectively, and $\Sigma$ is a Cauchy surface with future-directed unit normal vector $n^{\alpha}$. It can be shown that $\Omega(h,h')$ is independent of the choice of $\Sigma$ \cite{Fewster:2012bj,Friedman:1978wla}.  
If there were no degeneracy, i.e. if there were no solutions $h^{(null)}_{\mu\nu}$ satisfying $\Omega(h^{(null)},h)=0$ for all solutions $h_{\mu\nu}$, one could define an inner product by
\beq
\langle h,h' \rangle=i\Omega(\overline{h},h'), \label{innerproduct}
\eeq
where the overbar denotes complex conjugation. Suppose that a complete set of positive-frequency solutions, i.e. solutions whose time dependence is of the form $e^{-i \omega t}$, $\omega > 0$, is given by $\{h^{(n)}_{\mu\nu}\}$, where $n$ represents all (continuous and discrete) labels. Then a positive- and a negative-frequency solution would be orthogonal to one another with respect to the inner product (\ref{innerproduct}), and this inner product would be positive definite on the space of positive-frequency solutions. Then, we could expand the quantum field $\hat{h}_{\mu\nu}(x)$ as
\beq
\hat{h}_{\mu\nu}(x)=\sum_{n}\left[\hat{a}_{n}h_{\mu\nu}^{(n)}(x)+\hat{a}_n^{\dagger}\overline{h_{\mu\nu}^{(n)}}(x)\right]. \label{eq:quantumexpansion}
\eeq
The canonical equal-time commutation relations would be equivalent to
\beq
[\hat{a}_{m},\hat{a}_n^{\dagger}]=\left(M^{-1}\right)_{mn} \label{equaltime}
\eeq
and
\beq
[\hat{a}_{m},\hat{a}_n]=[\hat{a}_{m}^{\dagger},\hat{a}_n^{\dagger}]=0, \label{equaltime2}
\eeq
where $M^{-1}$ is the inverse of matrix $M^{mn}=\langle h^{(m)},h^{(n)} \rangle$. However, due to gauge invariance, the symplectic product given by (\ref{simplecticproduct}) is degenerate: a pure gauge solution $h^{(gauge)}_{\mu\nu}=\nabla_{\mu}\Lambda_{\nu}+\nabla_{\nu}\Lambda_{\mu}$ has vanishing symplectic product with any other solution, as it is well known. Thus, one needs to modify the quantization procedure described above. One standard way to proceed is to consider only the physical solutions, i.e. solutions satisfying gauge conditions that fix the gauge degrees of freedom completely. When all gauge degrees of freedom are eliminated, the symplectic product is non-degenerate and one quantizes the field by imposing the equal-time commutation relations given by 
Eqs.~(\ref{equaltime}) and (\ref{equaltime2}). We follow this procedure after fixing the gauge completely as described in 
Sec.~\ref{subsec:scalarandvectorpert} and normalizing the scalar- and vector-type modes, so that $M^{mn}=\delta^{mn}$, which may involve Dirac delta functions. Thus, we expand the quantum gravitational perturbation as in Eq.~(\ref{eq:quantumexpansion}) in terms of positive- and negative-frequency solutions given by 
Eqs.~(\ref{scalarmodes1})-(\ref{scalarmodes3}) and (\ref{vecmodes}), with definite frequencies $\omega$. We require the positive-frequency solutions to be normalized with respect to the inner product (\ref{innerproduct}) as follows:
\beqa
\langle h^{(P;\omega l m)},h^{(P';\omega' l' m')} \rangle=\delta^{PP'}\delta^{ll'}\delta^{mm'}\delta(\omega-\omega'), \nonumber \\
\label{normalization}
\eeqa
where $P=S,V$ labels the type of the perturbations, with $S$ denoting the scalar-type and $V$ denoting the vector-type perturbations. 

We write the positive-frequency modes of the master variables as
\beq
\Phi^{P}_{\omega l}(t,r)=e^{-i \omega t}u^{P}_{\omega l}(r), \ \ \omega >0.
\eeq
Then the functions $u^{P}_{\omega l}(r)$ satisfy the following Schr\"{o}dinger-like differential equation:
\beqa
-f(r)\frac{\diff}{\diff r}\left(f(r)\frac{\diff}{\diff r}u^{P}_{\omega l}(r)\right)+\left(V_P(r)-\omega^2\right)u^{P}_{\omega l}(r)=0. \nonumber \\
\label{schrodingerlike}
\eeqa
Close to and far away from the horizon, both effective potentials given by Eqs.~(\ref{scalarpotential}) and (\ref{vecpotential}) tend to zero. Hence the two independent solutions of Eq.~(\ref{schrodingerlike}) can be written as
\beq
u^{P,up}_{\omega l}\approx \begin{cases} A^{P}_{\omega l}(e^{i\omega r^{*}}+\mathcal{R}^{P, up}_{\omega l}e^{-i\omega r^{*}}), \hspace{0.25cm}r \gtrsim 2M,\\
A^{P}_{\omega l}\mathcal{T}^{P, up}_{\omega l}e^{i\omega r^{*}}, \hspace{0.5cm} r \gg 2M; \end{cases} \label{upboundaryconditions}
\eeq
\beq
u^{P,in}_{\omega l} \approx \begin{cases} A^{P}_{\omega l}\mathcal{T}^{P,in}_{\omega l}e^{-i\omega r^{*}}, \hspace{0.5cm}r \gtrsim 2M,\\
A^{P}_{\omega l}(e^{-i \omega r^{*}}+\mathcal{R}^{P,in}_{\omega l}e^{i \omega r^{*}}), \hspace{.25cm} r \gg 2M, \end{cases} \label{inboundaryconditions}
\eeq
where $r^{*}\equiv r+2M \log{\left(\frac{r}{2M}-1\right)}$ is the tortoise coordinate.

The modes $u^{P,up}_{\omega l}$ are purely incoming from the past horizon $\mathcal{H}^{-}$ while the modes $u^{P,in}_{\omega l}$ are purely incoming from the past null infinity $\mathcal{J}^{-}$. Using 
Eq.~(\ref{normalization}), we determine the asymptotic normalization constants $A^{P}_{\omega l}$ to be:
\beq
A^{V}_{\omega l}=\frac{1}{\sqrt{8 \pi \omega (l-1)(l+2)}} \label{eq:vecnormalconstant}
\eeq
and
\beq
A^{S}_{\omega l}=\frac{1}{\sqrt{2\pi \omega (l-1)l(l+1)(l+2)}}. \label{eq:scalarnormalconstant}
\eeq
We present the calculation of the inner product for the scalar-type modes, which is necessary for finding 
Eq.~(\ref{eq:scalarnormalconstant}), in Appendix \ref{app:scalarinner}.

\subsection{Gravitational radiation emission by a point particle} \label{subsec:gsremittedpower}
The point particle will contribute to the action with the interaction term given by
\beq
\hat{S}_{I}=\frac{\sqrt{32\pi}}{2}\int \diff^{4}x \sqrt{-g}T^{\mu\nu}(x)\hat{h}_{\mu\nu}(x),
\eeq 
where $T^{\mu\nu}$ is its energy-momentum tensor. Without loss of generality (due to the spherical symmetry of the problem), we consider the particle orbiting the black hole in the $\theta=\pi/2$ plane, at $r=R$, with angular velocity $\Omega$, as measured by a static asymptotic observer. Its $4$-velocity is written as 
\beq
u^{\mu}=(\gamma,0,0,\gamma\Omega),
\eeq
where 
\beq
\gamma=\frac{\diff t}{\diff\tau}=\frac{1}{\left[f(R)-R^2 \Omega^2\right]^{\frac{1}{2}}}.
\eeq
One can write the energy-momentum tensor as 
\beq
T^{\mu\nu}=\mu\frac{u^{\mu}u^{\nu}}{\gamma \sqrt{-g}}\delta(r-R)\delta(\theta-\pi/2)\delta(\varphi-\Omega t),
\eeq
where $\mu$ is the particle's mass.

We expand the graviton field $\hat{h}_{\mu\nu}(x)$ as:
\beqa
\hat{h}_{\mu\nu}(x)&=&\sum_{P,\lambda}\sum_{l=2}^{\infty}\sum_{m=-l}^{l}\int \limits_{0}^{\infty}\diff\omega\left[\hat{a}^{P,\lambda}_{lm}(\omega)h_{\mu\nu}^{(P,\lambda;\omega l m)}(x) \right.\nonumber \\
&&\left.+{{}\hat{a}^{P,\lambda}_{lm}}^{\dagger}(\omega)\overline{h_{\mu\nu}^{(P,\lambda;\omega l m)}}(x)\right]. \label{expansion}
\eeqa

To first order in perturbation theory, the emission amplitude of a $\lambda=in, up$ graviton of the $P$-type with quantum numbers $l,m$ and frequency $\omega$ is
\beq
\mathcal{A}^{P,\lambda;\omega l m}_{\mathrm{em}}=\langle P,\lambda;\omega l m|i\hat{S}_I|0 \rangle,  \label{amplitude}
\eeq 
which can be found to be
\beq
\mathcal{A}^{P,\lambda;\omega l m}_{\mathrm{em}}=\frac{i\sqrt{32\pi}}{2}\int \diff^{4}x\sqrt{-g}T^{\mu\nu}\overline{h_{\mu\nu}^{(P,\lambda;\omega l m)}}.
\eeq
Here, the initial state $|0\rangle$ is the one annihilated by all the $\hat{a}^{P,\lambda}_{lm}(\omega)$, i.e. the Boulware vacuum. If we had chosen the Unruh \cite{unruhvacuum} or Hartle-Hawking vacuum states \cite{Hartle:1976tp} (see also Refs. \cite{ISRAELvacuum, Kayreport}), then the transition rate calculated from the amplitude given by Eq.~(\ref{amplitude}) would be associated with the \textit{net} radiation emitted by the particle, since the absorption and stimulated emission rates (these two rates are induced by the thermal fluxes) give the same result.   
The emission amplitude $\mathcal{A}^{P,\lambda;\omega l m}_{\mathrm{em}}$ is proportional to $\delta(\omega-m\Omega)$, and hence the particle will only emit gravitons with the condition $\omega=m\Omega$ satisfied. In particular, since $\omega$ and $\Omega$ are both positive, only modes with $m\geq1$ will be emitted.

The emitted power for a graviton with a given type of labels $P,\lambda$ and quantum numbers $l,m$ reads
\beq
W^{P,\lambda;lm}_{\mathrm{em}}=\int\limits_{0}^{\infty}\diff\omega \ \omega \frac{|\mathcal{A}^{P,\lambda;\omega l m}_{\mathrm{em}}|^2}{T},
\eeq
where
\beq
T=2\pi\delta(0)=\int\limits_{-\infty}^{\infty}\diff t
\eeq
is the total time measured by an asymptotic static observer \cite{breuer1975gravitational}. Thus, the vector-type contributions to the emitted power are given by
\beqa
W^{V,\lambda;lm}_{\mathrm{em}}&=&64\pi^2\mu^2\gamma^2f(R)^2m\Omega^3\left|Y^{(lm)}_{\varphi}\left(\frac{\pi}{2},\Omega t\right)\right|^2 \nonumber \\ 
&& \times \left|\frac{\diff}{\diff R}\left(Ru^{V,\lambda}_{\omega_m l}(R)\right)\right|^2,  \label{vectorpower}
\eeqa
with
\beq
\omega_m=m\Omega.
\eeq
We note that the vector-type modes only contribute for odd values of $(l+m)$, due to the presence of the $\left|Y^{(lm)}_{\varphi}\left(\frac{\pi}{2},\Omega t\right)\right|^2$ factor, which vanishes for even values of $(l+m)$.
The scalar-type contributions can be written as
\beqa
W^{S,\lambda;lm}_{\mathrm{em}}&=&16\pi^2\mu^2\gamma^2m\Omega \left|Y^{lm}\left(\frac{\pi}{2},\Omega t\right)\right|^2 \nonumber \\
&&\times \left|F^{(\lambda;\omega_m l)}_{tt}(R)+2R^2\Omega^2F^{\lambda;\omega_m l}(R)\right|^2, \nonumber \\ \label{scalarpower}
\eeqa
where the functions $F^{(\lambda;\omega_m l)}_{tt}(R)$ and $F^{\lambda;\omega_m l}(R)$ are:
\beqa
F^{\lambda;\omega_m l}(R)&=&\frac{1}{4HR^2}\left\{[H-Rf'(R)]\left[RHu^{S,\lambda}_{\omega_m l}(R)\right] \right.\nonumber \\
&&\left. +2Rf(R)\frac{\diff}{\diff R}\left[RHu^{S,\lambda}_{\omega_m l}(R)\right]\right\}
\eeqa
and
\beqa
F^{(\lambda;\omega_m l)}_{tt}(R)&=&\frac{1}{2H}\left\{f(R)^2\frac{\diff^2}{\diff R^2}\left[RHu^{S,\lambda}_{\omega_m l}(R)\right]\right. \nonumber \\
&&\left. -\omega_m^2 RHu^{S,\lambda}_{\omega_m l}(R)\right\}.
\eeqa
We note that only the scalar-type modes with even values of $(l+m)$ contribute to the emitted power due to the $\left|Y^{lm}\left(\frac{\pi}{2},\Omega t\right)\right|^2$ factor.

Next, we compute numerically the solutions to Eqs.~(\ref{waveeqscalar}) and (\ref{waveeqvector}) in the frequency domain. We integrate numerically these differential equations by requiring the boundary conditions given by 
Eqs.~(\ref{upboundaryconditions}) and (\ref{inboundaryconditions}) to be satisfied, choosing suitable values of $r$. For the value close to the horizon, we have chosen $r/M \geq 2+\epsilon$, with $\epsilon=10^{-3}$. As for the ``numerical infinity'', $r_{\infty}$, we write it as a function of $l$ and $\omega_m$ such that the following condition is satisfied
\beq
\omega_m^2 \gg \frac{l(l+1)}{r_{\infty}^2}.
\eeq
In our computation we have chosen our ``numerical infinity'' to be:
\beq
r_{\infty}=\frac{250\sqrt{l(l+1)}}{\omega_m}.
\eeq

With the numerically obtained solutions, we use Eqs.~(\ref{vectorpower}) and  (\ref{scalarpower}) to compute the total emitted power as
\beq
W_{\mathrm{em}}= \sum\limits_{\lambda}\sum\limits_{l=2}^{\infty}\sum\limits_{m=1}^{l}\left(W^{S,\lambda;lm}_{\mathrm{em}}+W^{V,\lambda;lm}_{\mathrm{em}}\right). \label{totalpower}
\eeq
We compute the emitted power as a function of the angular velocity $\Omega$, relating to it the radial coordinate $R$ of the test particle by
\beq
R=\left(\frac{M}{\Omega^2}\right)^{1/3},
\eeq
which is required for the particle to be in circular orbit around the black hole, according to GR \cite{wald}. We include in our results both stable (up to $\Omega=(6\sqrt{6} M)^{-1}\approx 0.068~M^{-1}$) and unstable circular orbits (up to $\Omega=(3\sqrt{3} M)^{-1}\approx 0.192~M^{-1}$). As the circular orbit approaches the orbit of the light ray at $R=3M$, the emitted power increases rapidly, mainly because its motion becomes ultrarelativistic (the particle's energy increases with $u^t$). For this reason we find it more appropriate to plot the emitted power on a logarithmic scale. The particle becomes ultrarelativistic only for unstable orbits and we note that, if backreaction is taken into account, a particle in an unstable orbit starts its plunge into the black hole. 

\begin{figure}[!ht]
\includegraphics[scale=0.68]{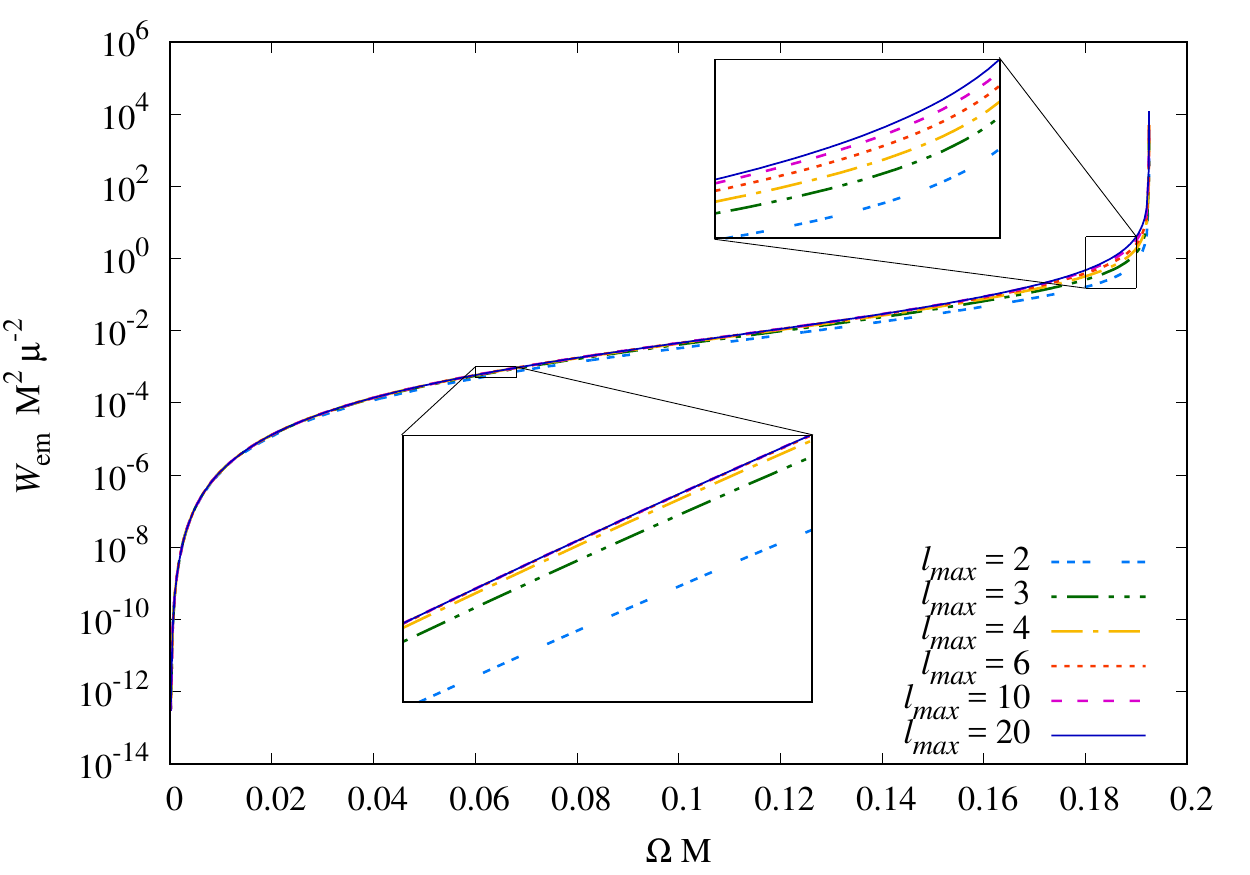}
\caption{\label{fig:totalpower} Total power emitted by the test particle rotating around the black hole, given by 
Eq.~(\ref{totalpower}), plotted as a function of the angular velocity $\Omega$. The summation in $l$ in 
Eq.~(\ref{totalpower}) is truncated at a certain value of $l$, which we denote as $l_{max}$.}
\end{figure}



\begin{figure}[ht]
\includegraphics[scale=0.68]{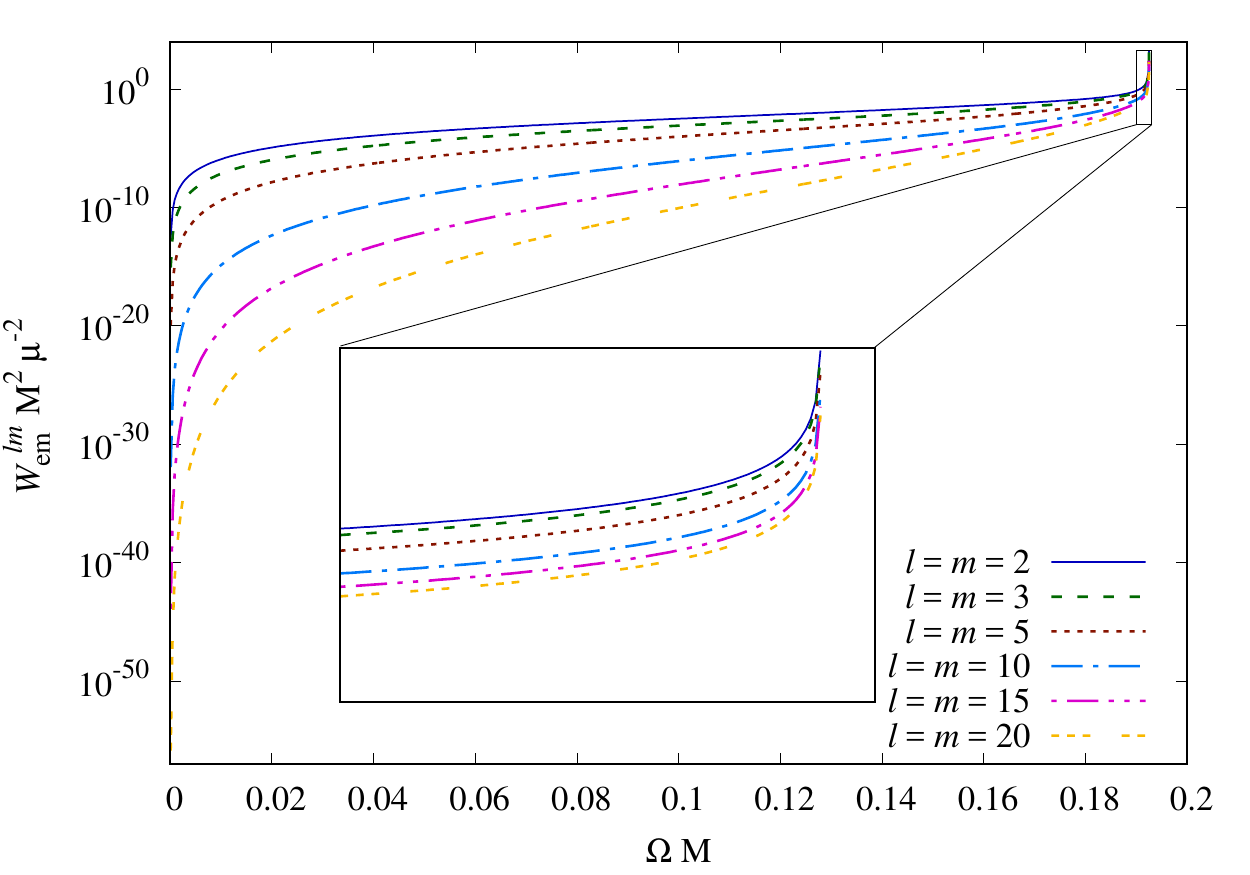}
\caption{\label{fig:scalarpower} Scalar-type power emitted by the test particle, given by Eq.~(\ref{scalarpower}), as a function of the angular velocity $\Omega$. We show here the modes with $l=m$, which give the main contributions to the total emitted power.}
\end{figure}

\begin{figure}[ht]
\includegraphics[scale=0.68]{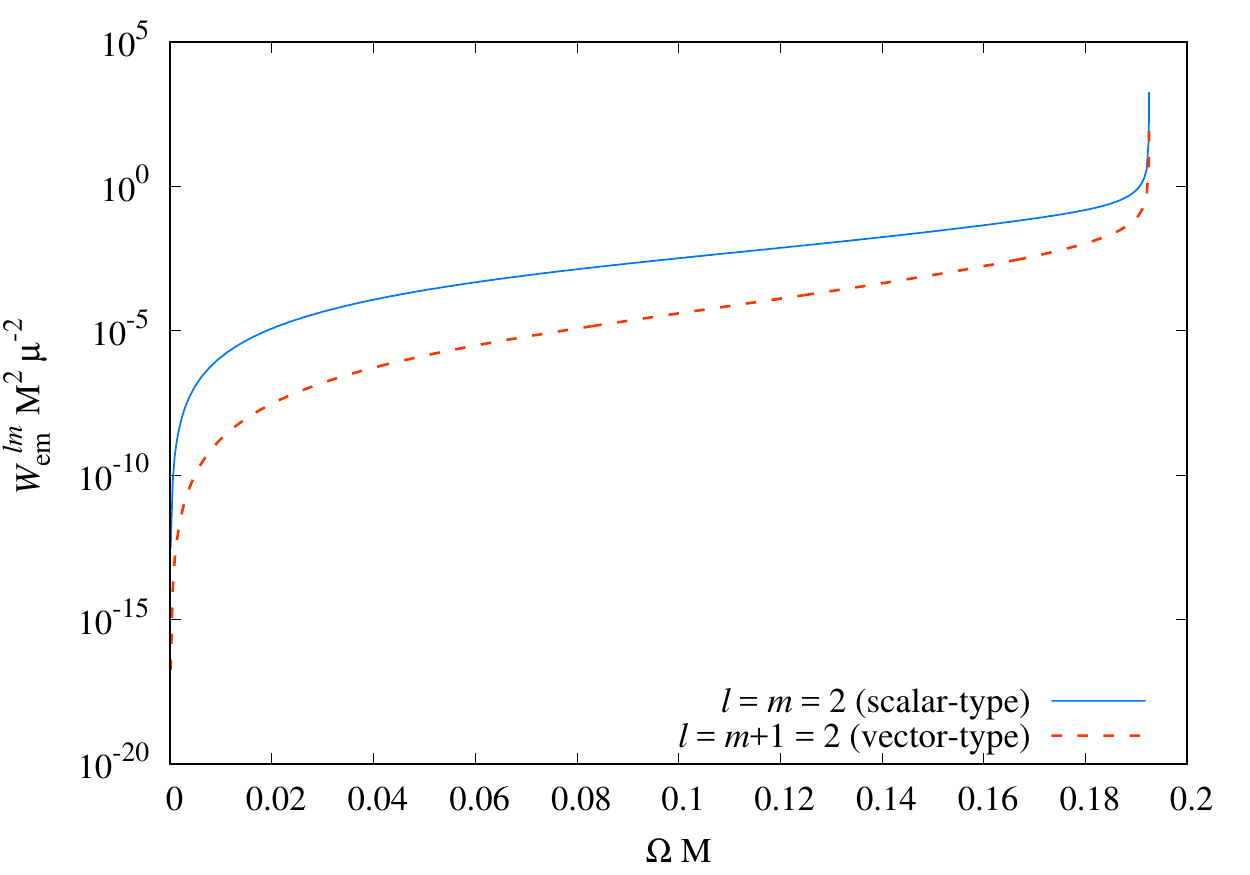}
\caption{\label{fig:vectorpower} Comparison between the $l=2$ scalar-type power, given by Eq.~(\ref{scalarpower}), and the $l=2$ vector-type power emitted by the test particle, given by Eq.~(\ref{vectorpower}), as a function of the angular velocity $\Omega$.}
\end{figure}

The results for the emitted power are plotted in Figs.~\ref{fig:totalpower}, \ref{fig:scalarpower} and \ref{fig:vectorpower}. For stable orbits, the main contribution to the total emitted power are the modes with $l=2$ ($m=2$ and $m=1$ for the scalar- and vector-type radiation, respectively). We note that the vector-type emitted power, given by Eq.~(\ref{vectorpower}), is suppressed by a factor of $R^2\Omega^2=M/R$, compared to the scalar-type emitted power \cite{poissongsr1}, given by Eq.~(\ref{scalarpower}) (see Fig.~\ref{fig:vectorpower}). 

\begin{figure}[!h]
\includegraphics[scale=0.68]{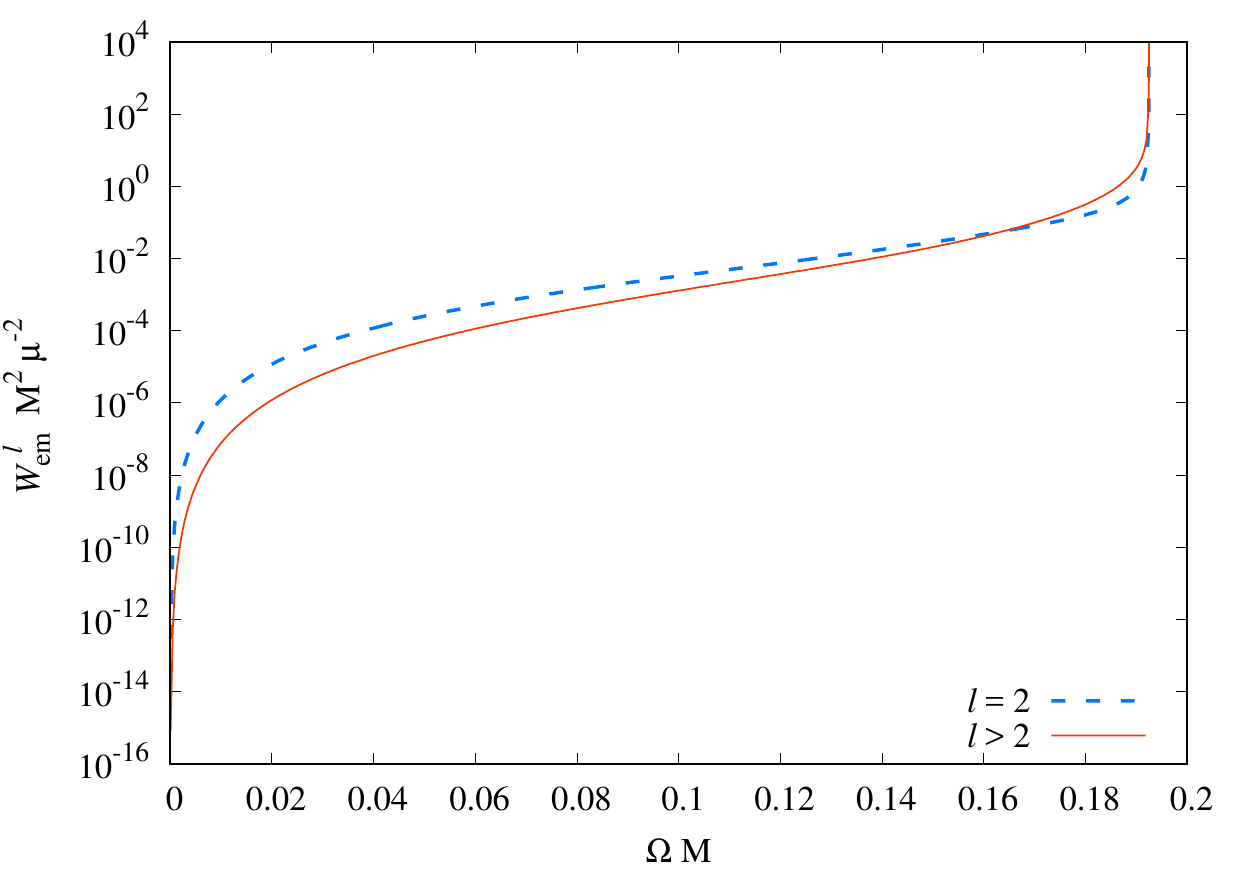}
\caption{\label{fig:scalarquadcomparison} Comparison between the contributions of the $l=2$ and $l>2$ modes (we considered contributions up to $l=20$). For most of the range of $\Omega$ (up to $\Omega \approx 0.164~M^{-1}$), the $l=2$ mode is the dominant contribution.}
\end{figure}

\begin{figure}[!h]
\includegraphics[scale=0.68]{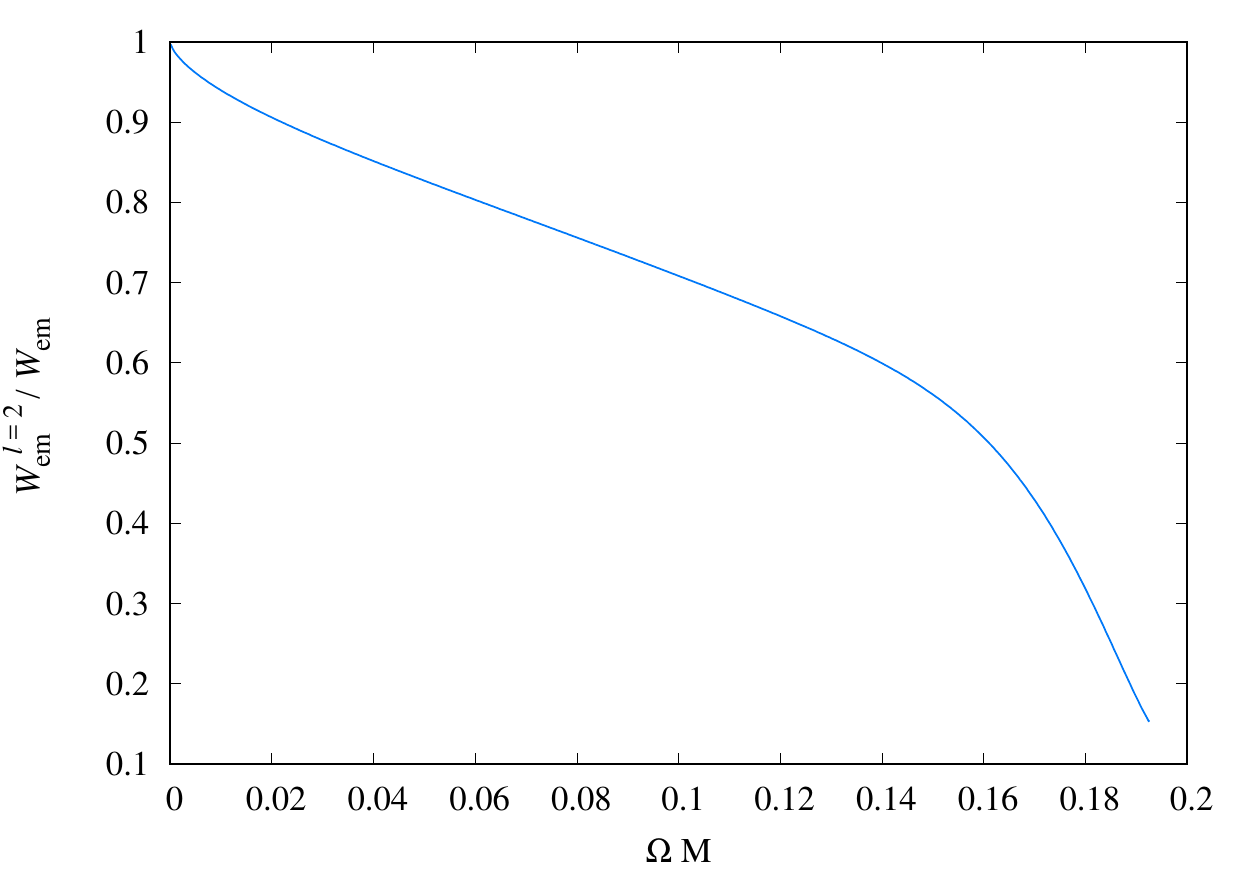}
\caption{\label{fig:ratiol2} Ratio $W^{l=2}_{\mathrm{em}}/W_{\mathrm{em}}$ between the emitted power of the $l=2$ modes and the total emitted power (we have considered contributions to the total emitted power up to $l=20$). The $l=2$ modes remain relevant contributions to the emitted power even for unstable orbits.}
\end{figure}

We note that the $l=2$ modes have a dominant contribution to the emitted power for most of the $\Omega$ range, as one can see in Fig. \ref{fig:scalarquadcomparison}. The contribution from all the other $l$ modes start to dominate over the $l=2$ mode contribution at $\Omega \approx 0.164~M^{-1}$.
For unstable orbits, the high-$l$ contributions become more important, but the $l=2$ modes still have a significant contribution (see Fig. \ref{fig:ratiol2}). 
For stable orbits, we can see that the contribution of the high multipoles is small. For unstable orbits, high multipoles are enhanced, but the low multipoles (especially the $l=2$ modes) are still important.

\begin{figure}
\includegraphics[scale=0.68]{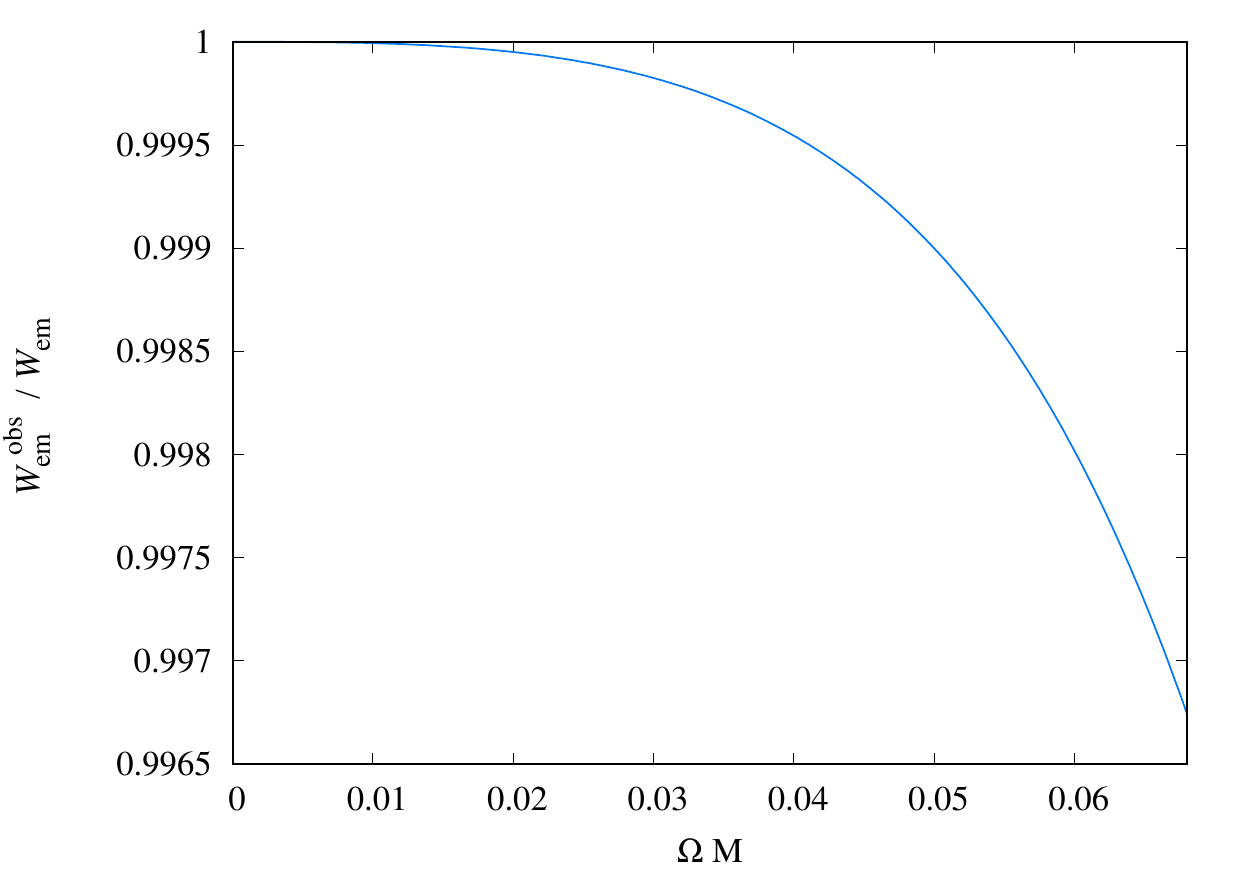}
\caption{\label{fig:obsratio} Ratio $W^{\mathrm{obs}}_{\mathrm{em}}/W_{\mathrm{em}}$, between the asymptotically observed and the total emitted power, as a function of $\Omega$, plotted for stable orbits. We have considered contributions up to $l=20$. We see that almost all the energy is radiated away to infinity, in the case of stable orbits.}
\end{figure}

\begin{figure}
\includegraphics[scale=0.68]{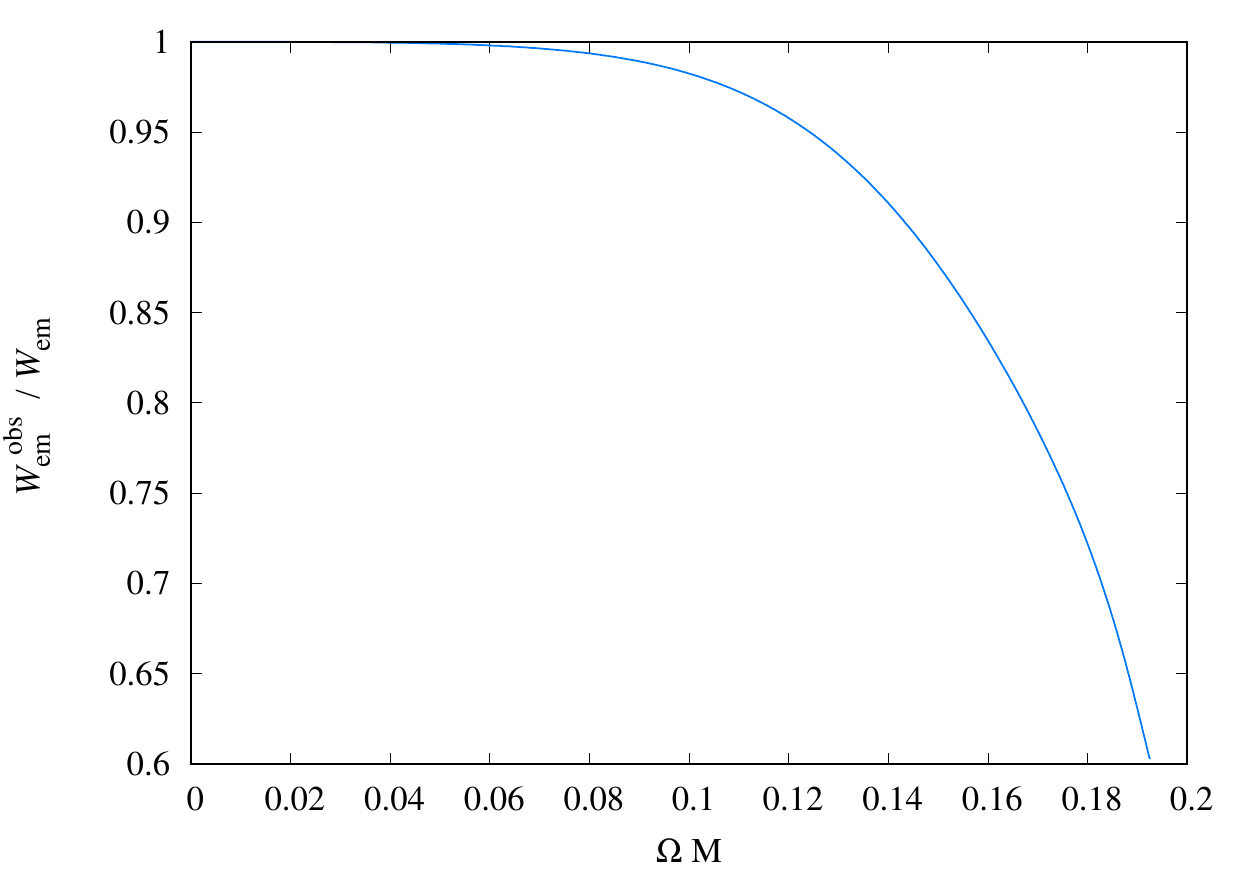}
\caption{\label{fig:obsratioall} Ratio $W^{\mathrm{obs}}_{\mathrm{em}}/W_{\mathrm{em}}$, between the asymptotically observed and the total emitted power, as a function of $\Omega$, plotted for stable, as well as for unstable circular orbits. As in Fig. \ref{fig:obsratio}, we have considered contributions up to $l=20$.}
\end{figure}

One can compute the power of emitted gravitons observed at infinity by considering only the modes which are purely outgoing at future null infinity. Since these modes are related to the modes $u^{P,in}_{\omega l}$ by complex conjugation, we can write the power observed at infinity as\footnote{The corresponding formulae for the massless scalar radiation in Ref. \cite{crispino1} [Eqs. (35) and (36) of that reference] are incorrect~\cite{ref:corrigendumcrispinocqg17}. Similar incorrect formulae were used in Refs. \cite{crispinoprd77,silva,crispinomacedo,meirafilho,crispino2}. Correction of these formulae does not affect the main conclusions in those references.}
\beqa
W^{\mathrm{obs}}_{\mathrm{em}}&=&\sum\limits_{P,l,m}W^{P,in;lm}_{\mathrm{em}}. \label{eq:observedpower}
\eeqa
In Figs. \ref{fig:obsratio} and \ref{fig:obsratioall}, the ratio $W^{\mathrm{obs}}_{\mathrm{em}}/W_{\mathrm{em}}$ is plotted as a function of the angular velocity $\Omega$. For unstable orbits, a considerable amount of emitted power is absorbed by the black hole, as shown in Fig. \ref{fig:obsratioall}. Approximately $38 \%$ of the radiation fails to reach the asymptotically flat region (infinity), for the innermost unstable orbit. In contrast, for stable orbits, more than $99 \%$ of the emitted power escapes to infinity, as one can see in Fig. \ref{fig:obsratio}.  We compared our results for the asymptotic radiation (scalar- and vector-type contributions) in stable orbits with other works \cite{poisson2, PhysRevD.7.1002}, resulting in excellent agreement.

\begin{figure}
\includegraphics[scale=0.68]{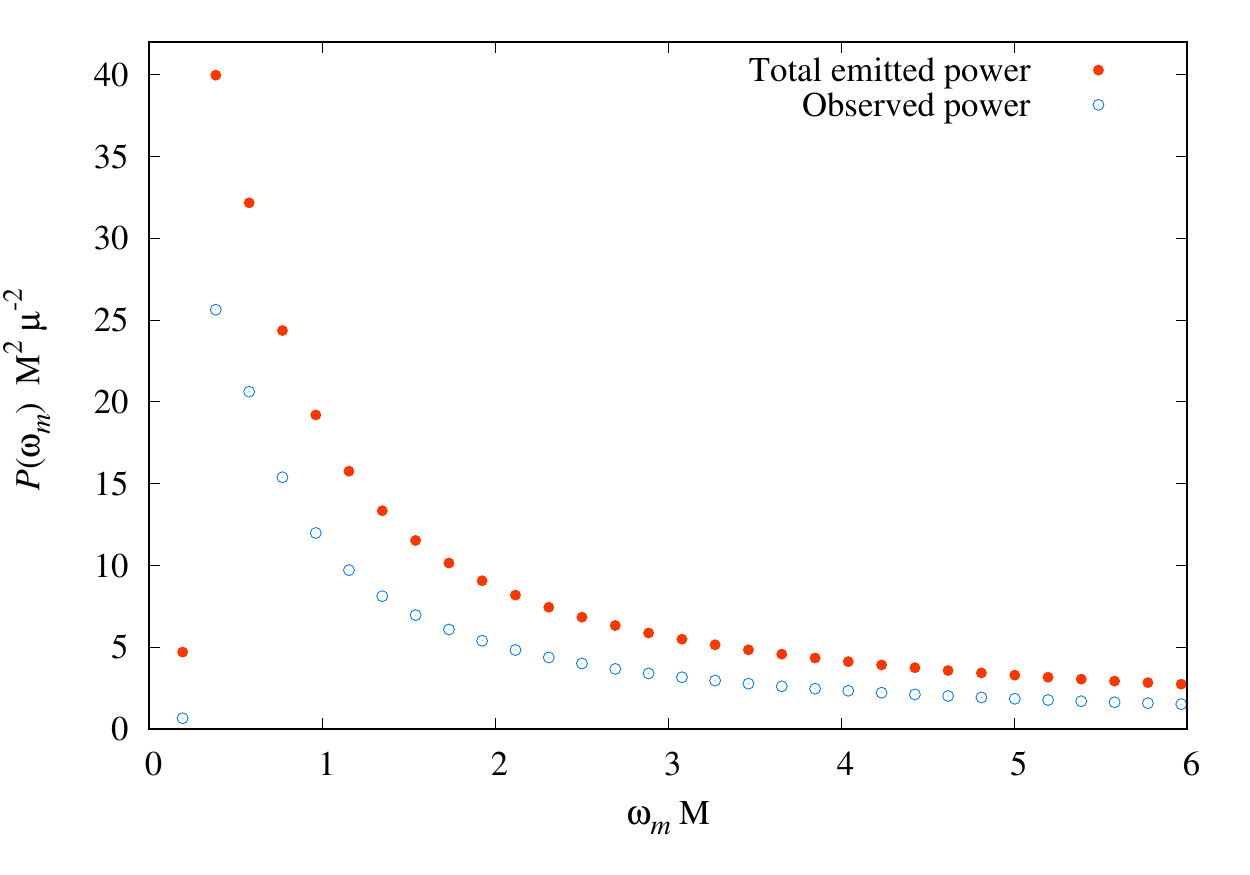}
\caption{\label{fig:lowspectrum} Total power emitted for a given frequency $\omega_m$. Frequencies up to $m=30$ are shown in this plot. Since only gravitons with frequencies that are integer multiples of the particle's angular velocity are emitted ($\omega_m= m\Omega$), the spectrum is discrete.}
\end{figure}

\begin{figure}
\includegraphics[scale=0.68]{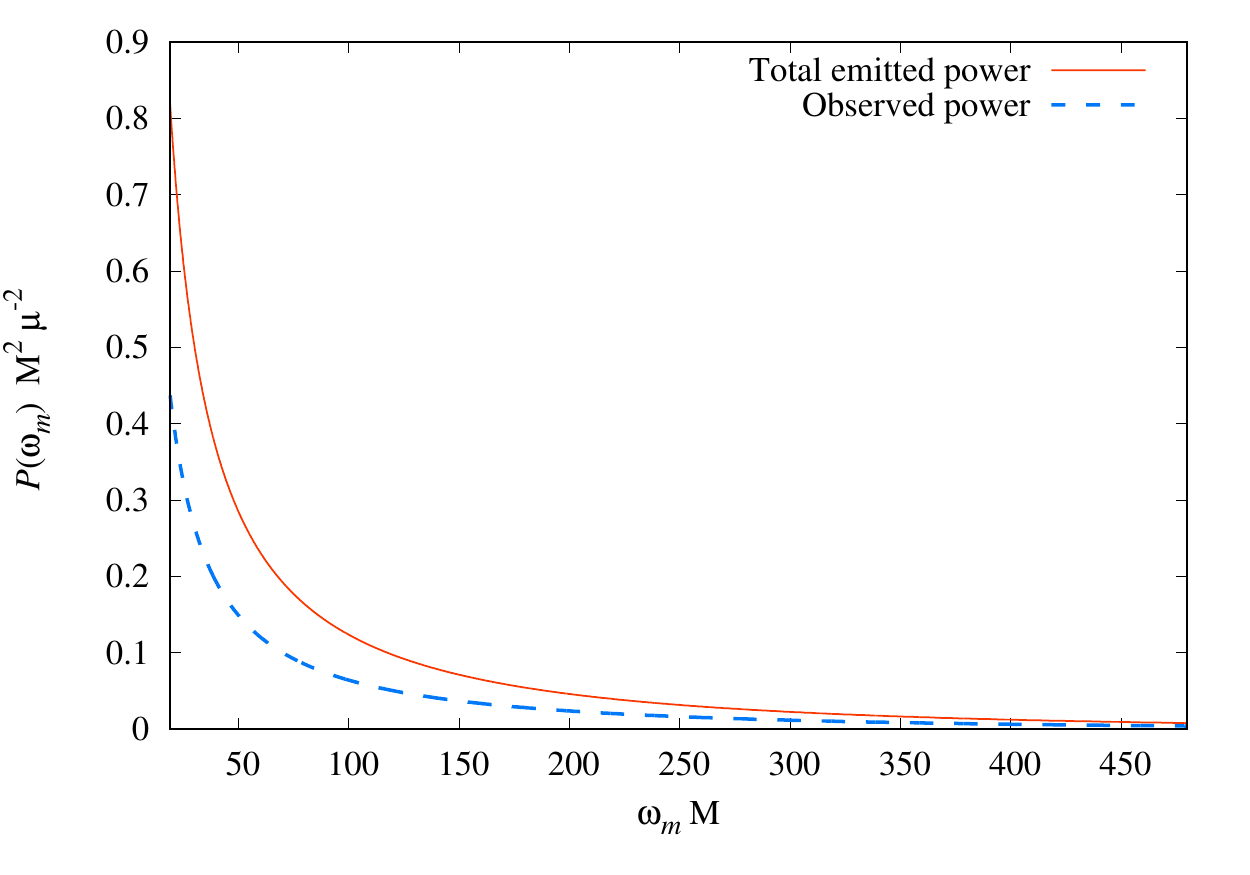}
\caption{\label{fig:highspectrum} Emitted power for a given frequency $\omega_m$ in the high-frequency range ($100 \leq m \leq 2500$).}
\end{figure}

Next, we analyze the radiation associated to the modes with large $m$, and hence with large $l$. Since only modes with $\omega_{m}=m \Omega$ are emitted, the total power for a given frequency, $P(\omega_m)$, can be written as a function of $m$, and hence as a function of $\omega_m$, as \cite{PhysRevD.7.1002}
\beq
P(\omega_m)=\sum\limits_{\lambda,P} \sum \limits_{l \geq |m|}^{\infty}W^{P,\lambda;lm}_{\mathrm{em}}. \label{spectrump}
\eeq
The power $P(\omega_m)$ depends on a discrete variable $\omega_m$, but we can regard it as a continuous variable for $m \gg 1$. In this continuum limit, we can write the emitted power $W_{\mathrm{em}}$ in terms of the spectral density function, denoted by $P(\omega)/\Omega$, as~\cite{PhysRevD.8.4309} 
\beq
W_{\mathrm{em}}=\sum\limits_{m=1}^{\infty}\frac{P(\omega_m)}{\Omega} \Delta \omega_m \approx \int\limits_{0}^{\infty} \diff \omega \frac{P(\omega)}{\Omega},
\eeq
with $\Delta \omega_m =\omega_m-\omega_{m-1}=\Omega$. 
We compute $P(\omega_m)$ and its asymptotically observed counterpart, obtained by summing only $W^{P,in;lm}_{\mathrm{em}}$ in Eq.~(\ref{spectrump}), for a particle orbiting the black hole in a highly relativistic unstable orbit with $R=(3+\delta)M$, $\delta=5\times10^{-4}$, for frequencies up to $\omega=2500\Omega$. We neglect the scalar-type modes with $l>m$ and vector-type modes with $l>m+1$ since the $l=m$ scalar-type and $l=m+1$ vector-type modes contribute to more than $99\%$ of the power at a given $l$. The results are shown in Figs.~\ref{fig:lowspectrum} and \ref{fig:highspectrum}. 
These results are in excellent agreement with those for high multipoles in Refs.~\cite{chitre, PhysRevD.7.1002} and show that the contribution of the low frequency modes is still relevant for the total radiation, even for unstable orbits. 

\section{Comparison with flat spacetime results} \label{sec:compareflat}

Let us now compare the emitted power $W_{\mathrm{em}}$ in Schwarzschild spacetime with its analogues in Minkowski spacetime, $W^{M}_{\mathrm{em}}$. For the flat spacetime computation, we consider the particle to be in a circular orbit bound to a stellar object, due to a Newtonian force. When the particle is not very close to the central object, the two cases should give similar results. 

In Minkowski spacetime, we use essentially the same procedure as in the Schwarzschild case to obtain the emitted power, with the difference that we set $f(r)=1$. The perturbed metric $h_{\mu\nu}$ will have the same form as in the Schwarzschild case, but now both master fields in Minkowski spacetime satisfy the same equation, namely
\beq
\Box \Phi^{M;P}_{l}(t,r)-V_{M}\Phi^{M;P}_{l}(t,r)=0, \label{minkowskieq}
\eeq
with $V_{M}=l(l+1)/r^2$ and $P=S,V$. The d'Alembertian in Eq.~(\ref{minkowskieq}) is the one compatible with the (flat) orbital spacetime line element $\diff s^2_{orb}=-\diff t^2+\diff r^2$. One can write positive-frequency solutions to Eq.~(\ref{minkowskieq}), which are regular at the origin, as
\beq
\Phi^{M;P}_{\omega l}(t,r)=C^{P}_{\omega l}e^{-i \omega t} r j_{l}(\omega r), \label{fieldsmink}
\eeq  
where $j_{l}(\omega r)$ are the spherical Bessel functions of the first kind \cite{abramowitz}.
Using the inner product given by Eq.~(\ref{innerproduct}), we obtain the normalization constants $C^{P}_{\omega l}$:
\beq
C^{V}_{\omega l}=\sqrt{\frac{\omega}{2 \pi (l-1)(l+2)}}
\eeq
and
\beq
C^{S}_{\omega l}=\sqrt{\frac{2\omega}{\pi (l-1)l(l+1)(l+2)}}.
\eeq

\begin{figure}[ht]
\includegraphics[scale=0.68]{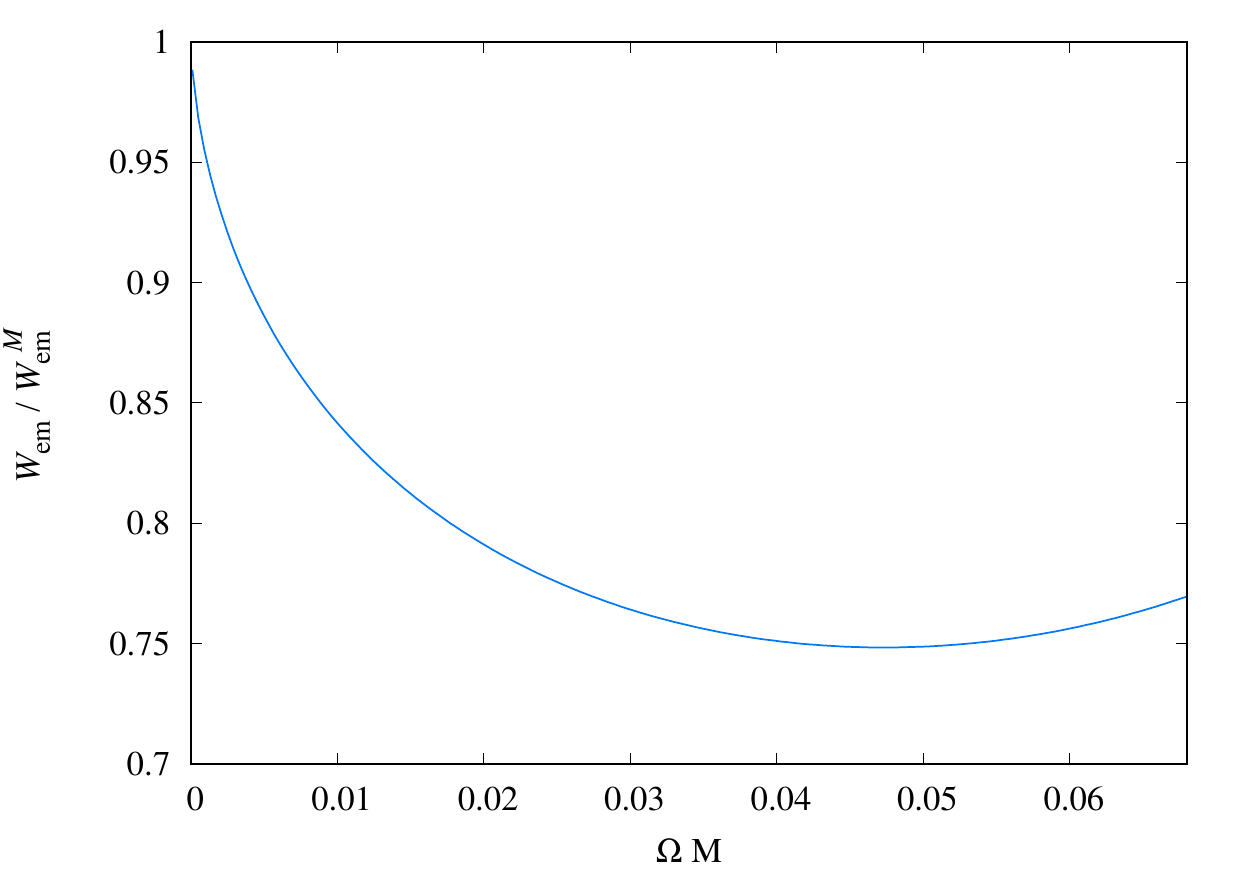}
\caption{\label{fig:wswm} Ratio $W_{\mathrm{em}}/W^{M}_{\mathrm{em}}$ as a function of $\Omega$. We have considered contributions up to $l=20$. The maximum value considered for $\Omega M$ is $(6\sqrt{6})^{-1}$. As $\Omega$ increases, the ratio decreases up to approximately $25 \%$, until it starts to increase as a result of the ultrarelativistic motion.}
\end{figure}

To compute the emitted power in Minkowski spacetime, we simply substitute the master fields of Eq.~(\ref{fieldsmink}) into Eqs. (\ref{vectorpower}) and (\ref{scalarpower}) to obtain\footnote{We adopt the usual Minkowski vacuum, i.e. the vacuum that is annihilated by the annihilation operators corresponding to the positive-frequency mode functions given by Eq.~(\ref{fieldsmink}).}
\beqa
W^{M;S;lm}_{\mathrm{em}}&=&16\pi^2\mu^2\gamma_{M}^2m\Omega \left|Y^{lm}\left(\frac{\pi}{2},\Omega t\right)\right|^2 \nonumber \\
&&\times \left|F^{(M;\omega_m l)}_{tt}(R_{M})+2R_{M}^2\Omega^2F^{M;\omega_m l}(R_M)\right|^2 \nonumber \\ \label{minkscalarpower}
\eeqa
and
\beqa
W^{M;V;lm}_{\mathrm{em}}&=&64\pi^2\mu^2\gamma_{M}^2m\Omega^3\left|Y^{(lm)}_{\varphi}\left(\frac{\pi}{2},\Omega t\right)\right|^2 \nonumber \\ 
&& \times \left|C^{V}_{\omega_m l}\frac{\diff}{\diff R_{M}}\left(R_{M}^2j_{l}(\omega_m R_M)\right)\right|^2, \nonumber \\ \label{minkvectorpower}
\eeqa
where $\gamma_{M}=(1-R_{M}^2 \Omega^2)^{-1/2}$ is the Lorentz factor. The quantities $F^{M;\omega_m l}(R_M)$ and $F^{(M;\omega_m l)}_{tt}(R_{M})$ are obtained by substituting the master field $\Phi^{M;S}_{\omega_m l}$ into Eqs. (\ref{F}) and (\ref{Fab}), respectively, with the mass of the black hole set to zero, namely
\beqa
F^{M;\omega_m l}(R_M)&=&\frac{C^{S}_{\omega_m l}}{4}\left\{ \vphantom{\frac{\diff^2}{\diff R_{M}^2}\left[R_M^2j_{l}(\omega_m R_M)\right]} \omega_m^2 R_M^2 j_{l}(\omega_m R_M) \right. \nonumber \\
&&+\left. \frac{\diff^2}{\diff R_{M}^2}\left[R_M^2j_{l}(\omega_m R_M)\right] \right\} \label{eq:minkowskiF}
\eeqa
and
\beqa
F^{(M;\omega_m l)}_{tt}(R_{M})&=&\frac{C^{S}_{\omega_m l}}{2}\left\{\frac{\diff^2}{\diff R_{M}^2}\left[R_M^2j_{l}(\omega_m R_M)\right] \right. \nonumber\\
&&\left.-\omega_m^2 R_M^2 j_{l}(\omega_m R_M) \vphantom{\frac{\diff^2}{\diff R_{M}^2}\left[R_M^2j_{l}(\omega_m R_M)\right]} \right\}.
\eeqa
We have used the equation of motion given by Eq.~(\ref{minkowskieq}) to simplify Eq.~(\ref{eq:minkowskiF}). In Newtonian gravity, for the particle to be in a circular orbit around the stellar object, its radial coordinate is related to its angular velocity by Kepler's third law, namely $R_M(\Omega)=(M\Omega^{-2})^{1/3}$. Since the angular velocity $\Omega$ is a quantity measured by an asymptotic static observer and, hence, coordinate independent, it is meaningful to compare the emitted powers from the orbiting particle in Schwarzschild and Minkowski spacetimes with the same value of $\Omega$. This comparison is plotted in Fig. \ref{fig:wswm}.

We note that there is a significant conceptual difference between the radiations in the two cases: in Schwarzschild spacetime, the circular orbit is a geodesics of the spacetime, whereas in Minkowski spacetime, it is a trajectory supported by an external force. In other words, the circular orbit is not a geodesic of Minkowski spacetime. Thus, the total radiation computation in Minkowski spacetime should include the radiation generated by the source of the external force, since the particle's energy-momentum tensor is not conserved by itself. However, it is likely that this additional radiation is negligible. (This is indeed the case if the circular motion is supported by a thin rod connecting the particle and the origin~\cite{PhysRevD.8.1640}.) The comparison in this section has been done primarily to show consistency of our numerical results in Schwarzschild spacetime by comparing them to the solutions obtained analitically in Minkowski spacetime, given by Eq.~(\ref{fieldsmink}), which should be a good approximation to the master fields far away from the black hole. 

\section{Concluding Remarks}
\label{sec:finalremarks}

We have computed the power of gravitational radiation emitted by a particle in circular orbit around a Schwarzschild black hole. By writing the gravitational perturbations in a gauge-invariant formalism, we used QFT at tree level to obtain numerically the emitted power, for both stable and unstable orbits. The scalar-type gravitational radiation was shown to be dominating over the vector-type gravitational radiation, the latter being suppressed by a factor of $R^{2}\Omega^2=M/R$. This result is in agreement with Ref. \cite{poissongsr1}, where the emitted power was computed for stable orbits only. For most of the range of the test particle's angular velocity, we have shown that the main contributions to the emitted power are from the $l=2$ modes for both scalar- and vector-type gravitational perturbations. In unstable orbits, the contributions from high multipole modes are enhanced. Nevertheless, the $l=2$ modes still have contributions that are far from negligible.      

In stable orbits, almost all of the emitted energy escapes to infinity. 
For unstable orbits however, a considerable amount of energy is absorbed by the black hole.

Comparing the emitted powers in Schwarzschild and Minkowski spacetimes, we see that the ratio between them (Schwarzschild over Minkowski) approaches unity when the particle is located very far away from the black hole. This ratio decreases as the orbit moves inwards but starts to increase near the innermost stable circular orbit ($R=6M$) because the orbit becomes ultrarelativistic sooner (i.e., for smaller values of $\Omega$) in the Schwarzschild spacetime, than in flat spacetime. 

We found that high multipoles of the gravitational radiation are significantly enhanced only for unstable orbits, although the low multipole modes remain as a relevant contribution to the emitted power even in this case. We also have found that no high frequency peaks are present in the power spectrum, unlike the scalar-radiation case~\cite{crispinoprd77}. 
We note that the particle in a stable orbit will emit radiation, gradually losing energy and inspiraling to orbits with smaller radii. When it reaches an unstable orbit, the particle emits a high amount of gravitational radiation and quickly plunges down into the black hole.

\begin{acknowledgments}
We would like to acknowledge Conselho Nacional de Desenvolvimento Cient\'ifico e Tecnol\'ogico (CNPq) and Coordena\c{c}\~ao de Aperfei\c{c}oamento de Pessoal de N\'ivel Superior (CAPES) for partial financial support.
A. H. also acknowledges partial support from the Abdus Salam International Centre for Theoretical Physics through the Visiting Scholar/Consultant Programme. A. H. thanks the Universidade Federal do Par\'a (UFPA) in Bel\'em and R. B. is grateful to the University of York for the kind hospitality while part of the work was carried out.

\end{acknowledgments}

\appendix
\section{Computation of the scalar-type inner product} \label{app:scalarinner}

In this Appendix, we compute the inner product for scalar-type perturbations at the future horizon, using Eddington-Finkelstein coordinates. For the scalar-type perturbation, the inner product (\ref{innerproduct}) reads
\beqa
\langle h^{S}, h'^{S} \rangle&=&-i \int\limits_{\Sigma}d\Sigma n_{a}J^{a}, \label{scalarinnerproduct}
\eeqa
where
\begin{widetext}
\beqa
J^{a}=\overline{Y^{lm}}Y^{l'm'}\left[\frac{4}{r}D^{d}r\left(\overline{F^{(\omega l)ab}}F^{(\omega' l')}_{bd}-F^{(\omega' l')ab}\overline{F^{(\omega l)}_{bd}}\right) -\left(\overline{F^{(\omega l)bc}}D^{a}F^{(\omega' l')}_{bc}-F^{(\omega' l')bc}D^{a}\overline{F^{(\omega l)}_{bc}}\right)\right], \label{innercurrent}
\eeqa
\end{widetext}
and $n_{a}$ is the (future-pointing) unit vector normal to the Cauchy hypersurface $\Sigma$. The inner product given by Eq.~(\ref{scalarinnerproduct}) can be rewritten as
\beqa
\langle h^{S}, h'^{S} \rangle&=&\frac{i\delta^{ll'}\delta^{mm'}(l-1)l(l+1)(l+2)}{2}\int\limits_{2M}^{\infty}\frac{\diff r}{f(r)} \nonumber \\
&& \times \left(\overline{\Phi^{S}_{\omega l}}\partial_t \Phi^{S}_{\omega' l'}-\Phi^{S}_{\omega' l'}\partial_t\overline{\Phi^{S}_{\omega l}}\right).  
 \label{simpinnerproductscalar}
\eeqa
We will derive Eq.~(\ref{simpinnerproductscalar}) using Eddington-Finkelstein coordinates. To simplify the notation, from now on we omit the labels for the frequency and angular quantum numbers, denoting quantities that depend on $\omega'$, $l'$ and $m'$ with a prime. 

Defining a new coordinate by
\beq
u \equiv t-r^{*},
\eeq
where $r^{*}$ is the tortoise coordinate, we find
\beq
\diff u= \diff t- \frac{\diff r}{f(r)}
\eeq
and the orbit spacetime line element (\ref{orbit}) becomes
\beqa
\diff s_{orb}^2&=& 
-f(r)\diff u^2-2\diff u \diff r.
\eeqa
We compute the inner product at the future horizon. The horizon is at $r=2M$ and $-\infty < u < \infty$. If $r < 2M$, the $r=$constant surface is a spacelike surface. A normalized (future-pointing) vector orthogonal to this surface can be written as
\beqa
n^{a}&=&-\left[-f(r)\right]^{-1/2}D^{a}r \nonumber \\
&=&\left[-f(r)\right]^{-1/2}\left(\frac{\partial}{\partial u}\right)^{a}+\left[-f(r)\right]^{1/2}\left(\frac{\partial}{\partial r}\right)^{a}. \label{eq:normalvec}
\eeqa
For this surface we have
\beq
\diff \Sigma=\diff \Omega_2 \diff u \left[-f(r)\right]^{1/2}r^2.
\eeq
Hence, using Eq.~(\ref{eq:normalvec}), we obtain
\beq
\diff \Sigma n^{a}=r^2 \diff \Omega_2 \diff u \left[\left(\frac{\partial}{\partial u}\right)^{a}-f(r)\left(\frac{\partial}{\partial r}\right)^{a}\right].
\eeq
In the limit $r \to 2M$, we get
\beq
\lim_{r \to 2M} \diff \Sigma n^{a}=4 M^2 \diff \Omega_2 \diff u \left(\frac{\partial}{\partial u}\right)^{a}.
\eeq
From one of the equations of motion ($g^{ab}F_{ab}=0$), we have
\beq
F_{ur}=\frac{f(r)}{2}F_{rr},
\eeq
which means that $F_{ur}$ vanishes at the horizon.
Moreover, the first term in Eq.~(\ref{innercurrent}) does not contribute to the inner product because
\beqa
n^{a}D^{d}r\left(\overline{{F_{a}}^{b}}F'_{bd}-{F'_{a}}^{b}\overline{F_{bd}}\right)=0.
\eeqa
This follows from the fact that $F_{ab}$ is a symmetric tensor and that $n^{a} \propto D^{a}r$ [see 
Eq.~(\ref{eq:normalvec})].

For the second term in Eq.~(\ref{innercurrent}), we have, at the horizon, 
\beqa
\overline{F^{bc}}D_{u}F'_{bc}-F'^{bc}D_{u}\overline{F_{bc}} &=& \overline{F_{rr}}\partial_u F'_{uu}+\overline{F_{uu}} \partial_u F'_{rr} \nonumber \\
&& - F'_{rr} \partial_u \overline{F_{uu}}-F'_{uu}\partial_u \overline{F_{rr}} \nonumber \\
&&+\frac{1}{M}\left(\overline{F_{rr}}F'_{uu}-F'_{rr}\overline{F_{uu}}\right). \nonumber \\ \label{secondterm}
\eeqa 
We integrate Eq.~(\ref{secondterm}) by parts with respect to $u$, indicating with the symbol ``$\approx$'' the equivalence under integration by parts. We find
\beqa
\overline{F^{bc}}D_{u}F'_{bc}-F'^{bc}D_{u}\overline{F_{bc}} &\approx&2\left(\overline{F_{rr}}\partial_u F'_{uu}+\overline{F_{uu}} \partial_u F'_{rr}\right) \nonumber \\
&&+\frac{1}{M}\left(\overline{F_{rr}}F'_{uu}-F'_{rr}\overline{F_{uu}}\right).  \nonumber \\
\label{efinnercurrent}
\eeqa
At the horizon, one can write the components of the gauge-invariant quantity $F_{ab}$ as
\beq
F_{uu}=\left(2M\partial^2_{u}+\frac{1}{2}\partial_u\right)\Phi^S \label{Fuu}
\eeq
and
\beq
F_{rr}=\left[2M\partial^2_{r}+\left(2-\frac{6}{H}\right)\partial_r\right]\Phi^S. \label{Frr}
\eeq
Here, $H=H(r=2M)$ [see Eq.~(\ref{eq:hfunction})]. Thus,
\beq
H=l^2+l+1.
\eeq
We note that we set $r=2M$ after all differentiation is done.
Now, we substitute Eqs.~(\ref{Fuu}) and (\ref{Frr}) into Eq.~(\ref{efinnercurrent}), 
and integrate by parts with respect to $u$ to obtain
\beqa
\overline{F^{bc}}D_{u}F'_{bc}-F'^{bc}D_{u}\overline{F_{bc}}&\approx&\partial_u \Phi'^{S}\hat{O}\overline{\Phi^{S}}-\partial_u\overline{\Phi^{S}} \hat{O} \Phi'^{S}, \nonumber \\
\eeqa 
where the fourth order differential operator $\hat{O}$ reads
\beqa
\hat{O}&=&8M^2\partial_u^2 \partial_r^2-6M\partial_u\partial^2_r+8M\left(1-\frac{3}{H}\right)\partial_u^2\partial_r \nonumber\\
&& -6\left(1-\frac{3}{H}\right)\partial_u \partial_r+\partial_r^2+\frac{1}{M}\left(1-\frac{3}{H}\right)\partial_r. \nonumber \\
\label{4do}
\eeqa
We write $\hat{O}$ as the following linear combination
\beq
\hat{O}=A\Box\left[\frac{f(r)}{V_S(r)}\Box\right]+(B+C\partial_u)\Box, \label{linear4do}
\eeq
with $A$, $B$ and $C$ being suitably chosen constants. We can use the equations of motion to write
\beq
\hat{O}\Phi^S_{\omega l}=\underbrace{\frac{V_S(r)}{f(r)}[(A+B)}_{\text{constant at} \ r=2M}\Phi^S_{\omega l}+C\partial_u\Phi^S_{\omega l}].
\eeq
We note that the term containing $C$ does not contribute to the inner product. However, its presence is needed to write the fourth order operator (\ref{4do}) in the form (\ref{linear4do}). We may use a symbolic computation software to obtain
\beq
A=\frac{[3+(l-1)l(l+1)(l+2)]^2}{2[3+l(l+1)(l^4+2l^3-l+1)]},
\eeq
\beqa
B=2-\frac{3}{l^2+l+1}-\frac{3}{2}\frac{l^2+l+1}{(l-1)l(l+1)(l+2)+3}, \nonumber \\
\eeqa
and
\beq
C=6M\frac{l^2+l+1}{(l-1)l(l+1)(l+2)+3}.
\eeq

Then, at the horizon, the inner product (\ref{scalarinnerproduct}) can be written as
\begin{widetext}
\beqa
\langle h^{S}, h'^{S} \rangle&=&i\frac{(l-1)l(l+1)(l+2)}{2}\int \diff \Omega_2 \diff u \overline{Y^{lm}}Y^{l' m'} \left(\overline{\Phi^S_{\omega l}}\partial_u\Phi^S_{\omega' l'}-\Phi^S_{\omega' l'}\partial_u\overline{\Phi^S_{\omega l}}\right) \nonumber \\
&=&\lim_{r \to 2M} i\frac{(l-1)l(l+1)(l+2)}{2} \int\limits_{\Sigma} \diff \Sigma n^{a}\overline{Y^{lm}}Y^{l' m'}\left(\overline{\Phi^S_{\omega l}}\partial_a\Phi^S_{\omega' l'}-\Phi^S_{\omega' l'}\partial_a\overline{\Phi^S_{\omega l}}\right).
\eeqa
This can be evaluated in a $t=$constant Cauchy surface in $tr$ coordinates as
\beqa
\langle h^{(S;\omega l m)}, h^{(S;\omega' l' m')} \rangle&=&i\frac{(l-1)l(l+1)(l+2)}{2} \int\limits_{S^2}\diff \Omega_2 \overline{Y^{lm}}Y^{l' m'}  \int\limits_{2M}^{\infty}  \frac{\diff r}{f(r)} \left(\overline{\Phi^S_{\omega l}}\partial_t\Phi^S_{\omega' l'}-\Phi^S_{\omega' l'}\partial_t\overline{\Phi^S_{\omega l}}\right),
\eeqa
which leads to Eq.~(\ref{simpinnerproductscalar}).
\end{widetext}
Another way to obtain Eq.~(\ref{simpinnerproductscalar}) is by computing the inner product (\ref{scalarinnerproduct}) directly using $tr$ coordinates. However, this method is much more involved, requiring several cumbersome integration by parts, although we can still use a computational software to do all the algebraic computations. We have done so and the same result as the Eddington-Finkelstein one has been obtained, as expected. In addition, since computing the inner product in these coordinates does not require the presence of a horizon, we can also use this method in flat spacetime, in spherical coordinates.

%
\end{document}